\documentclass[a4paper,11pt]{article}
\usepackage{jheppub}
\usepackage{eurosym}
\usepackage{amssymb}
\usepackage{amsfonts,cite}
\usepackage{amsmath}
\usepackage{graphicx}
\usepackage{multirow}
\usepackage{xcolor}
\usepackage{epsfig}
\usepackage{fancyhdr}

\newbox\mybox

\newcommand\fverb{\setbox\mybox=\hbox\bgroup\verb}
\newcommand\fverbdo{\egroup\medskip\noindent\fbox{\unhbox\mybox}\ }
\newcommand\fverbit{\egroup\item[\fbox{\unhbox\mybox}]}

\abstract{We study the quantum Pais-Uhlenbeck oscillator at the resonant
	(equal-frequency) point, where the dynamics becomes non-diagonalisable and the
	conventional Fock-space construction collapses. At the classical level, the
	degenerate system admits more than one Hamiltonian formulation generating the
	same equations of motion, leading to a nontrivial quantisation ambiguity.
	Working first in the ghostly two-dimensional Hamiltonian formulation, we construct differential
	intertwiners that generate a spectrum-generating algebra acting on the
	generalised eigenspaces of the Hamiltonian. This algebra organises the
	generalised eigenvectors into finite Jordan chains and closes into a hidden
	$su(2)$ Lie algebra that exists only at resonance.
	
	We then show that quantising a classically equivalent Hamiltonian yields a
	radically different quantum theory, with a fully diagonalisable spectrum and
	genuine degeneracies. Our results demonstrate that the resonant
	Pais-Uhlenbeck oscillator provides a concrete example in which classically
	equivalent Hamiltonians define inequivalent quantum theories.}  
	
	\author[a]{Andreas Fring,}
	\author[b]{Ian Marquette,}
	\author[c]{Takano Taira}

\title{Spectrum-generating algebra and intertwiners of the resonant Pais-Uhlenbeck oscillator}

\affiliation[a]{Department of Mathematics, City St George's, University of London,  Northampton Square, \\ London EC1V 0HB, UK}
\affiliation[b]{Department of Mathematical and Physical Sciences, La Trobe University, Bendigo, VIC 3552, Australia}
\affiliation[c]{Department of Physics, Kyushu University
	744 Motooka, Nishi-ku, Fukuoka City, 819-0395, Japan}
	
\emailAdd{a.fring@city.ac.uk}	
\emailAdd{i.marquette@latrobe.edu.au}
\emailAdd{taira.takano.292@m.kyushu-u.ac.jp}	
	
\begin{document}
	\maketitle
	
	\pagestyle{fancy}
	\fancyhead{} 
	\fancyhead[LE,RO]{\small\itshape  Spectrum-generating algebra and intertwiners of the resonant Pais-Uhlenbeck oscillator} 
	
	\renewcommand{\headrulewidth}{0.4pt}
	
\section{Introduction}	

Higher time derivative theories (HTDTs) arise naturally in many areas of theoretical physics, including effective field theory, where higher derivative operators appear as suppressed corrections at low energies \cite{buchbinder2017effective}; modified gravity, in which curvature-squared and more general higher derivative actions play a central role \cite{stelle77ren,starobinsky1980new,sotiriou2010f}; regularisation schemes designed to improve ultraviolet behaviour \cite{stelle77ren,grav3}; and attempts at ultraviolet completion of quantum field theories based on higher derivative or nonlocal dynamics \cite{tomboulis2015renormalization,modesto2012super,modesto2016superrenormalizable,barnaby2008dynamics}.

Despite their appealing features, such as improved renormalisability and richer dynamical structure, they are notoriously difficult to quantise consistently. The fundamental obstruction is the appearance of Ostrogradsky instabilities \cite{ghostconst,motohashi1,motohashi4}, which manifest themselves quantum mechanically through Hamiltonians that are unbounded from below or through states with negative or indefinite norm \cite{weldon03quant,Woodard1,fring2024higher,fring2025quant}. Understanding whether, and in what sense, such theories can admit a meaningful quantum interpretation therefore remains an important open problem.

The Pais–Uhlenbeck (PU) oscillator \cite{pais1950field,bolonek2005ham,mann2005dirac,dam2006,bolonek2006comment,bender2008no,smilga2009comments,most2010h,mosta2011im,andrzejewski2014ham,Sugg3,smilga2017class,mandal2022bfv,elbistan2023various} occupies a distinguished position in this context. As the simplest mechanical system governed by a higher order equation of motion, it serves as a prototype for HTDTs. In its standard form the PU model describes a fourth-order oscillator characterised by two frequencies $\omega_1$ and $\omega_2$, and it can be rewritten as a two-dimensional system with an indefinite kinetic term in various ways \cite{FFT}. This reformulation exposes its “ghostly’’ nature \cite{diez2024foundations}: depending on the parameter regime, one encounters either normalisable states with spectra unbounded from below or non-normalisable states with bounded spectra \cite{fring2025quant}. 

In recent work, the non-degenerate PU model $\omega_1 \neq \omega_2$ has been shown to admit several remarkable algebraic and geometric structures. In particular, Lie-algebraic symmetries, intertwining operators, and bi-Hamiltonian and tri-Hamiltonian formulations have been exploited to construct ghost-free representations and to identify parameter regimes in which the dynamics can be generated by positive-definite Hamiltonians \cite{FFT,felski2026three}. These developments have improved our understanding of how the usual Ostrogradsky pathology may be circumvented, at least in certain non-degenerate settings.

By contrast, the degenerate limit $\omega_1 =\omega_2=: \omega$ remains much less well understood. This limit is far from innocuous: it corresponds to a coalescence of the two oscillatory modes and is accompanied by profound structural changes in both the classical and quantum theory. At the classical level, the system develops additional constants of motion and is no longer diagonalisable into two independent harmonic oscillators. The oscillatory solutions become resonant\footnote{We emphasise that the term ``resonant'' is used here in the classical
	sense of linear differential equations: it refers to the coincidence of normal
	frequencies and the resulting appearance of secular terms and Jordan blocks in
	the time-evolution operator. This notion of resonance should not be confused
	with quantum resonances associated with metastable states and complex energy
	poles in scattering theory.}, so that the general solution contains secular terms proportional to $t \cos(\omega t)$ and $t \sin(\omega t)$.  At the quantum level, the standard Fock-space construction breaks down, ladder operators collapse, and Jordan-block-type features are known to emerge. In this limit, only two of the four quantum sectors of the
	non-degenerate model survive, and precisely those whose wavefunctions are
	non-normalisable; the sectors admitting normalisable states disappear entirely
	at resonance. One may still attempt to apply the same scheme and obtain the degenerate case by a limiting procedure \cite{mann2005dirac}, but, as was argued in \cite{bolonek2006comment}, this limit is ill-defined. The subsequent analysis in \cite{kaparulin2020resonance} once more confirms that the degenerate model is structurally distinct from the non-degenerate one; interestingly, it was also shown there that the inclusion of certain interactions can lead to dynamics that is bounded from below, although a fully consistent quantised version has not yet been provided. For these reasons, the degenerate PU oscillator remains a challenge as a particularly delicate and singular point in the parameter space of HTDTs.

The purpose of this paper is to carry out a systematic analysis of the quantum
degenerate Pais-Uhlenbeck model from several complementary perspectives, with
particular emphasis on its algebraic organisation and on the ambiguities of
quantisation that arise at resonance. At the classical level, the resonant
system admits more than one Hamiltonian formulation generating exactly the same
equations of motion, raising the fundamental question of which Hamiltonian
should be quantised.

We work primarily with the two-dimensional “ghostly” Hamiltonian formulation
and impose the specific coupling that enforces the resonance condition. We
first show how the usual supersymmetric/Darboux intertwining construction
survives in a drastically simplified form at this point, even though
the associated eigenfunctions are no longer square integrable. These
intertwiners naturally lead to the emergence of a spectrum-generating algebra
that acts on the generalised eigenspaces of the Hamiltonian and organises the
spectrum into finite Jordan chains, despite the collapse of the conventional
Fock-space picture. We show that, in the present model, this algebra closes into
a hidden $su(2)$ Lie algebra that exists only at resonance.

We then investigate how different choices of classically equivalent
Hamiltonians lead to inequivalent quantum theories. In particular, while the
ghostly Hamiltonian gives rise to a non-diagonalisable quantum theory organised
by Jordan chains, an alternative Hamiltonian formulation yields a fully
diagonalisable theory with genuine degeneracies. This demonstrates explicitly
that, in higher-derivative systems, classical equivalence does not guarantee
quantum equivalence, and that the choice of Hamiltonian is an essential part of
the quantisation problem.

Next, we investigate whether the powerful bi-Hamiltonian mechanism that proves
successful in the non-degenerate model can still be used to construct a
positive-definite Hamiltonian at the resonant point. We show that this
possibility fails in a decisive way: although a second Hamiltonian and Poisson
tensor exist, no linear combination of them can be made positive definite in
any parameter regime. This establishes a sharp qualitative distinction between
the degenerate and non-degenerate PU models.

Finally, we introduce a factorisation of the ground-state wavefunction that
isolates one square-integrable direction in a restricted parameter window and
leads to effective one-dimensional Hamiltonians. This provides a further
perspective on how partial ghost-free behaviour may survive even at the
degenerate point, albeit in a limited and highly constrained sense.

Taken together, these results show that the degenerate PU oscillator is not
merely a limiting case of the generic theory, but a genuinely distinct
dynamical system with its own spectrum-generating algebra, algebraic
obstructions, and quantisation subtleties. As such, it provides a particularly
suitable arena for testing proposed resolutions of the ghost problem in HTDTs and
for probing the limits of algebraic and geometric quantisation methods in HTDTs.

Our paper is organised as follows: In section 2 we briefly review the classical degenerate PU oscillator, emphasising the resonant structure of the
equations of motion and the associated Jordan-type behaviour of the
time-evolution operator. In section 3 we analyse the quantum theory based on the
ghostly two-dimensional Hamiltonian formulation and construct a system of
intertwining operators that generates a hidden spectrum-generating $su(2)$
algebra organising the non-diagonalisable spectrum into finite Jordan chains.
In section 4 discuss a classically equivalent Hamiltonian
formulation and that was shown in section 3 to possess as quantisation that leads to a radically different,
fully diagonalisable quantum theory with genuine degeneracies, thereby
demonstrating the inequivalence of different quantisations at resonance. In
section 5 we explore a partial factorisation of the formal ground state and
investigate whether an effective one-dimensional, partially normalisable sector
can be isolated. Section 6 contains our conclusions and an outlook on possible
ghost-free quantisation schemes.

\section{The resonant point in the ghostly two-dimensional formulation }
We recall from \cite{fring2025quant,FFT} that the standard higher time derivative  PU-Hamiltonian 
\begin{equation}
	H_{PU}(q,\dot{q},\ddot{q},\dddot{q}) = \frac{1}{2} \ddot{q}^2- \frac{1}{2}( \omega_1^2 +  \omega_2^2  )\dot{q}^2 - \frac{1}{2} \omega_1^2   \omega_2^2 q^2 - \dot{q} \dddot{q} \qquad \omega_1, \omega_2 \in \mathbb{R} , \label{PUmodel}
\end{equation}	
transforms into the two dimensional model
\begin{equation}
	H_g(x,y,p_x,p_y) =    p_x^2 - p_y^2   + \nu^2 x^2+  \Omega  y^2  + g x y, \qquad \nu,\Omega,g   \in \mathbb{R} , \label{Hghost}
\end{equation}
with the linear transformations
\begin{eqnarray}
	q  &\rightarrow &-\frac{x+y}{2 \sqrt{2} \sqrt{\nu ^2+\Omega -g }}, \quad
	\dot{q}  \rightarrow   -\frac{p_x+p_y}{\sqrt{2} \sqrt{\nu ^2+\Omega -g }} ,   \quad
	\ddot{q}   \rightarrow  \frac{\left(2 \nu ^2-g\right)x+(g-2 \Omega ) y }{\sqrt{2} \sqrt{\nu ^2+\Omega-g} }, \qquad \label{qtoxy1} \\
		\dddot{q}   &\rightarrow & \sqrt{2} \frac{\left(2 \nu ^2-g\right)p_x+(g-2 \Omega ) p_y }{ \sqrt{\nu ^2+\Omega-g} }, \quad \omega_{1,2}= \sqrt{2} \sqrt{\nu ^2-\Omega \mp \sqrt{\left(\nu ^2+\Omega \right)^2-g^2}}.  \label{qtoxy}
\end{eqnarray}
Due to the Lorentzian signature of the kinetic term, this model is referred to as {\em ghostly}, since its quantum version possesses sectors with normalisable states but spectra unbounded from below, as well as non-normalisable states with bounded spectra. These characteristics reflect precisely the key features of HTDTs.

Here we consider the model for the special choice of the coupling constant
\begin{equation}
	g \;\rightarrow\; -(\nu^2+\Omega),
	\qquad \Rightarrow \qquad \omega_1=\omega_2=\omega .
	\label{degcoupling}
\end{equation}
This choice is in fact unique within the present formulation. Although the
alternative sign, $g=+(\nu^2+\Omega)$, also formally leads to equal frequencies,
the canonical transformation to the ghostly two-dimensional Hamiltonian then ceases to be valid. For this admissible choice of coupling, the classical dynamics reduces, in terms
of the original configuration variable $q(t)$, to the fourth-order equation of
motion of the resonant (degenerate) PU oscillator
\begin{equation}
	\left(\frac{d^2}{dt^2}+\omega^2\right)^2 q(t)=0 . \label{PUeom}
\end{equation}
The general solution to this equation contains secular terms proportional to
$t\cos(\omega t)$ and $t\sin(\omega t)$,
reflecting the coalescence of the two oscillatory modes at $\omega_1=\omega_2$. Equivalently, when the dynamics is
rewritten as a first-order linear system on phase space, $\dot X = A X$, with $X = \{ q,\dot{q},\ddot{q},\dddot{q} \}$ and a $4\times4$ constant matrix $A$, whose characteristic
polynomial is $(\lambda^2+\omega^2)^2$. The eigenvalues $\lambda=\pm i\omega$
thus have algebraic multiplicity two, but $A$ admits only one eigenvector for
each eigenvalue. Consequently $A$ is not diagonalisable and is similar to the
direct sum of two $2\times2$ Jordan blocks, i.e. $A=U J U^{-1}$ with $J = J_+ \oplus J_-$,
\begin{equation}
	J_\pm=\begin{pmatrix}\pm i\omega & 1 \\ 0 & \pm i\omega\end{pmatrix}=\pm i\omega\,\mathbf{I}+N, \quad \text{with}\,\, N^2=0 .
\end{equation}
The nontrivial nilpotent part $N$ is responsible for the secular terms in the
classical solution. At the quantum level, this manifests itself in the fact that the Hamiltonian, when restricted to a degenerate energy sector, decomposes as $H = E \mathbf{I} + N$ with a nontrivial nilpotent part $N$, so that time evolution and operator actions necessarily involve Jordan chains. This in turn motivates the search for ladder operators and a
spectrum-generating algebra acting within these sectors, which we construct
below.

In the ghostly two-dimensional formulation, Hamilton’s equations generated by
$H_g$ in (\ref{Hghost}) at the resonant point read
\begin{align}
	\dot x &= \frac{\partial H_g}{\partial p_x}=2p_x,  &	\dot y&= \frac{\partial H_g}{\partial p_y}=-2p_y, \label{flow1} \\
	\dot p_x &= -\frac{\partial H_g}{\partial x}
	= -2\nu^2 x + (\nu^2+\Omega)y,    &	   \dot p_y&= -\frac{\partial H_g}{\partial y}
	= -2\Omega y + (\nu^2+\Omega)x . \label{flow2}
\end{align}
We may combine this system into two coupled second order equations as
\begin{equation}
	\ddot x = -4\nu^2 x + 2(\nu^2+\Omega)y, \qquad  \ddot y = 4\Omega y - 2(\nu^2+\Omega)x ,
\end{equation}
which are equivalent to (\ref{PUeom}), when using the inverse transformation of (\ref{qtoxy1}) and (\ref{qtoxy}).  

In \cite{fring2025quant} the theory was fully quantised and the ``ground state'' with corresponding eigenvalue equation was identified as
\begin{equation}
	\psi_0(x,y) = c_0 e^{-\frac{\alpha  x^2}{2}-\frac{\beta  y^2}{2} +\gamma  x y   } , \qquad   \text{with} \quad H_g	\psi_0(x,y) =(\alpha -\beta)  	\psi_0(x,y) ,\label{wansatz}
\end{equation}
with $c_0$ denoting some overall normalisation constant. When expressing the auxiliary constants $\alpha$, $\beta$, $\gamma$ in terms of the model parameters $\nu$, $\Omega$, $g$, four solutions labelled by $\epsilon=\pm 1$ and $\eta=\pm 1$ were identified
\begin{equation}
	\alpha_\epsilon^\eta = \frac{2 \nu ^2  + \sigma_\epsilon }{ \Sigma_\epsilon^\eta}, \quad
	\beta_\epsilon^\eta = \frac{2 \Omega - \sigma_\epsilon}{\Sigma_\epsilon^\eta}, \quad
	\gamma_\epsilon^\eta = \frac{-g}{ \Sigma_\epsilon^\eta }, \,\, \Sigma_\epsilon^\eta = 2 \eta \sqrt{ \nu ^2-\Omega+ \sigma_\epsilon },
	\,\, \sigma_\epsilon =  \epsilon \sqrt{g^2-4 \nu ^2 \Omega } .  \label{abcd}
\end{equation}
For the choice of the coupling constant as in (\ref{degcoupling}), the constants  $\alpha$, $\beta$, $\gamma$ become singular or complex when $\nu^2 - \Omega> 0$ or  $\nu^2 - \Omega< 0$, respectively. Thus only the two ($\epsilon =1$)-sectors remain with the additional constraint $ \nu ^2-\Omega>0$, for which the constants simplify to
\begin{equation}
	\alpha^\eta =  \eta \frac{3 \nu ^2-\Omega }{ 2 \sqrt{2}  \sqrt{\nu ^2-\Omega }}  , \quad
	\beta^\eta =  - \eta \frac{\nu ^2-3 \Omega }{  2 \sqrt{2} \sqrt{\nu ^2-\Omega }}   , \quad
	\gamma^\eta =   \eta \frac{\nu ^2+\Omega }{  2 \sqrt{2} \sqrt{\nu ^2-\Omega }}      .  \label{abcdsimp}
\end{equation}
As we expect, the wavefunction $\psi_0(x,y)$ is not square integrable as the normalisation conditions
\begin{equation}
	\alpha^\eta >0, \qquad \beta^\eta >0, \qquad \alpha \beta - \gamma^2 = \frac{\Omega - \nu^2}{2}>0,
\end{equation}
can not be satisfied. However, for $\eta=1$ the spectrum is bounded from below which makes it worth exploring this sector further.

\subsection{Intertwining operators}

Next we carry out a supersymmetric/Darboux transformation  \cite{Witten:1981nf,Cooper:1982dm,bagchi2000super}/\cite{darboux,crum,matveevdarboux}. Defining the Hamiltonian 
\begin{equation}
	H_1  =    H_g -\alpha+\beta ,
\end{equation}
with the standard identification $p_x = - i \partial_x$, $p_y = - i \partial_y$ in mind, we find the right and left first order intertwining operators
\begin{equation}
A_- = (\alpha -\gamma) \partial_x + (\gamma-\beta) \partial_y + \frac{1}{2} (\nu^2 - \Omega) (x-y), \quad
A_+ = \frac{1}{2} (\partial_x +\partial_y )  +\frac{\alpha-\gamma}{2}(y-x),
\end{equation}
satisfying the relations
\begin{equation}
	A_- H_1 = H_g A_-, \qquad \text{and} \qquad   H_1   A_+ = A_+ H_g ,
\end{equation}
together with $[A_+,A_-]=0$.

Noting that $A_- \psi_0 =0$ the standard procedure yields the eigenvalue equations
\begin{equation}
	H_1   \psi_n  = n(\alpha - \beta)  \psi_n ,  \quad \text{and} \quad
    H_2   \psi_n = (n+1)(\alpha - \beta)  \psi_n, \quad  \text{with} \,\, \psi_n =A_+^n \psi_0  .
\end{equation}
None of the wavefunctions $\psi_n$ is square integrable. Acting with $A_+^n$ on $\psi_n$ will produce expressions of the form $\psi_n = P_n (x,y) \psi_0 $ with $ P_n $ being $n$-th order polynomials in $x,y$. In \cite{fring2025quant}  generic polynomials for all values of g were constructed, which precisely reduce to the expressions obtained here. We have 
\begin{equation}
 P_0 =1, \quad  P_1 \sim (x-y), \quad   P_2 \sim (x-y)^2 \,\, \ldots \,\, P_n = -\eta  2^{-n/2} \left(\nu ^2-\Omega \right)^{n/2} (x-y)^n .   \label{poly}
\end{equation}
Notice that the standard bosonic Fock space construction presented in \cite{fring2025quant} is not valid at the degenerate point, which is to be expected as it was constructed from the oscillatory solution of the non-degenerate scenario. However, we can utilise a hidden $su(2)$ Lie algebraic structure to solve the eigenvalue problem.

\section{A hidden spectrum-generating $su(2)$ algebra}

In \cite{cannata2010exactly,marquette22lad1,marquette22lad2}, a non-Hermitian model similar to our
ghostly system in~(\ref{Hghost}) was investigated. The difference between the
two models is that in \cite{marquette22lad1,marquette22lad2} has a well-defined kinetic term
with a Euclidean metric but a purely complex coupling constant. There, hidden
algebraic structures similar to those known for the two-dimensional real
harmonic oscillator were identified.

In our setting here, the resonant and non-diagonalisable nature of the degenerate
dynamics entails the appearance of a nontrivial nilpotent operator (associated
with the Jordan-type structure of the time-evolution operator) that mixes
generalised eigenvectors belonging to the same eigenvalue. In contrast to the
usual role of an $su(2)$ symmetry, which produces multiplets of degenerate
eigenstates, the present algebra organises finite Jordan chains associated with
a single eigenvalue. This naturally motivates the search for a
spectrum-generating algebra acting on these chains, which we now uncover in the
form of a hidden $su(2)$ symmetry. We refer to this algebra as {\em hidden} because it is not manifest in the original Hamiltonian or canonical variables, but emerges only after analysing the intertwining structure induced by the resonant Jordan-block dynamics.

 For this purpose we define new operators closely related to the intertwining operators
\begin{equation}
	a^\pm = \frac{1}{2}  \left[  \partial_x +\partial_y  \pm   \kappa    (y-x)    \right], \quad
	b^\pm = \frac{1}{2}  \left[  \partial_x -\partial_y  \pm \frac{\sqrt{2}}{\kappa}  \left(  \Omega y -  \nu^2 x   \right)    \right],
\end{equation}
with constants 
\begin{equation}
 \kappa:=  \frac{1}{\sqrt{2}} \sqrt{\nu^2 - \Omega}, \qquad     \lambda :=  \frac{1}{\sqrt{2}} \frac{\nu^2 + \Omega}{\ \sqrt{\nu^2 - \Omega}  },
\end{equation}
obeying the commutation relations
\begin{equation}
	\left[ a^+ , a^-   \right]= 	\left[a^+ , b^+   \right]=0, \qquad
	\left[ b^\pm  , a^\mp    \right]= \pm \kappa ,   \qquad 
		\left[ b^+ , b^-   \right]=  \lambda.
\end{equation}
These operators serve as well as intertwining operators and quasi-intertwining operators for $H_g$
\begin{equation}
	\left[  H_g, a^\pm   \right]   = \pm 2 \kappa \, a^\pm, \qquad 
	\left[  H_g, b^\pm   \right]   = \pm 2 \kappa \, b^\pm  \pm 2 \lambda \, a^\pm .
\end{equation}
Next we introduce the new operators
\begin{equation}
   R:= a^+ a^-, \quad S:= b^+ b^-, \quad T^\pm := a^+ b^-   \pm b^+ a^-, 
\end{equation}
obeying the mutual commutation relations
\begin{eqnarray}
	\left[ T^+, T^-   \right] &=& 2 \lambda R, \quad  
	\left[ R, S   \right] = - \kappa T^-, \quad  
	\left[ R, T^-   \right] = 2 \kappa R, \quad
		\left[ S, T^-   \right] =  \lambda T^+ - 2 \kappa S ,\qquad \\
	\left[T^+, R   \right] &=&	\left[ H_g, R   \right] = 		\left[ H_g, T^+   \right] =0, \quad  
	\left[ H_g, S   \right] =  2 \lambda T^-, \quad
		\left[ H_g, T^-  \right] =  - 4 \lambda R. \quad
\end{eqnarray}
Motivated by the structure of the commutation relations among $R$, $S$, $T^\pm$, we introduce the operators
\begin{equation}
	M_0 := \frac{1}{2 \kappa}  T^-, \qquad 
	M_- :=- \frac{2 \lambda}{\kappa}  R, \qquad 
	M_+ :=   \frac{1}{2 \kappa}   \left( \frac{1}{2 \kappa}  T^+  - \frac{1}{\lambda} S - \frac{\lambda}{ 4 \kappa^2} R  \right) .
\end{equation}
A direct computation then shows that these generators close a $su(2)$ Lie algebra
\begin{equation}
        \left[ M_0 , M_\pm \right]  = \pm M_\pm, \qquad    \left[    M_+ , M_-   \right]= 2 M_0 .
\end{equation}
We also define the operator
\begin{equation}
    K := \frac{\lambda }{ \kappa^2} R -\frac{1}{\kappa} T^+ + 1
\end{equation}
that lies in the center of the representation as it commutes with all generators
\begin{equation}
     \left[K , M_0     \right] =0, \qquad  \left[K , M_\pm     \right] =0.
\end{equation}
It is related to the standard $su(2)$-Casimir operator $C$ as 
\begin{equation}
	C= M_0^2+ \frac{1}{2} \left(M_+ M_- + M_- M_+    \right)= \frac{1}{4} \left( K^2 -1    \right),
\end{equation}
and to our Hamiltonian as 
\begin{equation}
	H_g = 2 \kappa K + M_- .
\end{equation}
Our explicit representation of this algebra in terms of differential operators acting on the space of test functions $f(x,y)$ reads
\begin{eqnarray}
	H_g &=& \partial_y^2-\partial_x^2+\nu ^2 x^2 +y^2 \Omega  -x y \left(\nu ^2+\Omega \right),  \\
	K  &=&  \frac{1}{2 \sqrt{2} \left(\nu ^2-\Omega \right)^{3/2}} \left[  \left(3\nu ^2-\Omega \right) \partial_y^2+2 \left(\nu ^2+\Omega \right) \partial_x \partial_y - \left(\nu ^2-3 \Omega \right) \partial_x^2  \right] \notag  \\	
	&&  +      \frac{1}{4 \sqrt{2} \left(\nu ^2-\Omega \right)^{1/2}}  \left[ \nu ^2 (3 x+y)-\Omega  (x+3 y)\right]  (x-y)  ,\\
	M_0 &=&  \frac{1}{4 \left(\nu ^2-\Omega \right)}  \left\{   \left[\nu ^2 (x+y)+\Omega  (x-3 y)\right]  \partial_x  +  \left[\nu ^2 (3 x-y)-\Omega  (x+y)\right] \partial_y   \right\} , \\
	M_- &=&  \frac{ \nu ^2+\Omega   }{4
		\left(\nu ^2-\Omega \right)}    \left[  -2 \left(\partial_y^2+2 \partial_x \partial_y +\partial_x^2\right) + \left(\nu ^2-\Omega \right) (x-y)^2   \right] , \\
	M_+ &=&  \frac{1}{32 (\nu ^4-\Omega^2 )}   \left\{ \left[\nu ^2 (3 x+y)-\Omega  (x+3 y)\right]^2 - 2  \left[   \frac{\nu^2 -3 \Omega}{\nu ^2-\Omega } \partial_x +
	   \frac{\Omega - 3 \nu^2}{\nu ^2-\Omega } \partial_y          \right]^2 \right\}. \qquad \,\,
	\end{eqnarray}
	Note that $M_-$ defines a positive quadratic form on the natural test-function domain when $\nu^2 > \Omega$. More related directly to the PU-model, using (\ref{qtoxy}), we may also represent these differential operators as acting on the space of test functions $f(q, \ddot{q})$. Defining 
	\begin{equation}
		D_n^\pm := \partial_q \pm n (\nu^2 - \Omega) \partial_{\ddot{q}}, \qquad
		q_n^\pm := \ddot{q} \pm n (\nu^2 - \Omega) q ,
	\end{equation}
	we have
	\begin{eqnarray}
		H_g &=&  D_2^- \partial_{\ddot{q}}  +\frac{1}{2} q_2^+ q_2^-,   \\
			K & =&   \frac{1}{4 \sqrt{2} \sqrt{\nu^2 - \Omega}}  \left[ \frac{1}{2 (\nu^2 -\Omega)} D_6^+ D_2^-   + q_6^- q_2^+   \right], \\
		M_0 &=&  -\frac{1}{4} \left[   \frac{1}{2 (\nu^2 -\Omega)} q_2^- \partial_q  +  q_6^+  \partial_{\ddot{q}}     \right],  \\
			M_- &= &  -\frac{1}{4} \left[   \frac{1}{2 (\nu^2 -\Omega)} (D_2^-)^2  -  (q_2^+)^2      \right],\\ 
		M_+ &=&  -\frac{1}{32 (\nu^2 -\Omega)} \left[   \frac{1}{2 (\nu^2 -\Omega)} (D_6^+)^2  -   (q_6^-)^2      \right] . 
	\end{eqnarray}
	
	Since $[H_g , K] =0$ and $	H_g = 2\kappa K + M_- $, the Hamiltonian preserves each fixed-$K$ eigenspace and may be analysed
	independently on every sector
	\begin{equation}
	\mathcal H_k := \{ \psi \;|\; K\psi = k\,\psi \}, \qquad k\in\mathbb N_0 .
	\end{equation}
	On such a sector we have
	\begin{equation}
	     H_g \psi = (2 \kappa k) \psi  + M_- \psi.
	\end{equation}
	It follows immediately that any vector $\psi \in \mathcal H_k$ and $ \psi \in ker M_-$, i.e. satisfying  $M_- \psi =0$, is a ``formal'' eigenstate of $H_g$. These lowest-weight
	vectors therefore provide the genuine eigenstates of $H_g$ in each $K$-sector. Solving the first-order equation $M_- \psi_m(x,y)=0$ yields the family
		\begin{equation}
 \psi_m(x,y) = (x-y)^m   e^{ - \frac{\alpha}{2}  x^2  - \frac{\beta}{2}  y^2 + \gamma x y  }, \quad m \in \mathbb{N}_0.
	\end{equation}
	which satisfy
		\begin{eqnarray}
		K  \psi_{k-1} &=& k \psi_{k-1} , \quad  \\
		   M_0  \psi_{k} &=& - \frac{k}{2} \psi_{k}, \quad   \\ 
		M_+  \psi_{k} &=&  \frac{k}{\nu^2 + \Omega}  \left[ \chi \psi_{k-1}  + (1-k)  \psi_{k-2}   \right],
		\qquad   \chi:= \frac{\nu ^2 (3 x+y)-\Omega  (x+3 y)}{ 2 \sqrt{2} \sqrt{\nu^2 - \Omega} } ,  \qquad \\
		M_+^2  \psi_{k} &=&  \frac{k(k-1)}{(\nu^2 + \Omega)^2}  \left[ \chi^2 \psi_{k-2}  +2  \chi  (2-k)  \psi_{k-3}   + (3-k)(2-k)   \psi_{k-4}   \right], \\
		& \vdots & \notag \\
		M_+^\ell  \psi_{k} &=&  \frac{1}{(\nu^2 + \Omega)^\ell }  \sum_{j=0}^\ell   \left(  -1 \right)^j     \binom{n}{k}      \chi^{n-j}        \frac{k!}{(k-n-j)! }  \psi_{k-n-j} .
	\end{eqnarray}
		Thus, repeated application of the raising operator $M_+$ generates a finite chain
		\begin{equation}
	\psi_{k,\ell} := M_+^\ell \psi_{k-1}  \in \mathcal H_k , \qquad \ell=0,1,\dots,k-1 ,
	   	\end{equation}
	which terminates after $k$ steps,
	\begin{equation}
		M_+^{k}\,\psi_{k-1} = 0 .
	\end{equation}
	For $\ell\ge 1$, however, the states $\psi_{k,\ell}$ are not eigenstates of
	$H_g$. Rather, they form a Jordan chain associated with the eigenvalue
	$2\kappa k$, in the sense that
	\begin{equation}
		(H_g - 2\kappa k)\,\psi_{k,0} = 0, \qquad
		(H_g - 2\kappa k)\,\psi_{k,\ell} = \ell(k-\ell) \psi_{k,\ell-1}, \quad \ell\ge 1 .
	\end{equation}
	The explicit coefficient $\ell(k-\ell)$ follows from the $su(2)$ commutation
	relations and ensures the termination of the chain at $\ell=k$.
	Thus each fixed-$K$ sector carries a finite-dimensional indecomposable
	representation of $su(2)$, whose states organise the non-diagonalisable
	(Jordan-type) action of $H_g$ at resonance.
	
	Writing $k=n+1$, the genuine eigenvalues of $H_g$ are therefore
	\begin{equation}
		E_n = 2\kappa (n+1) = (n+1)(\alpha-\beta), \qquad n=0,1,2,\dots
	\end{equation}
	with corresponding (formal) eigenstates $\psi_n=\psi_{k-1}$.

	This agrees with the findings of \cite{fring2025quant}, see equation (\ref{poly}). Thus, while the eigenspectrum is bounded from below in this sector, we stress that for real $\nu^2>\Omega$ these wavefunctions are not normalisable in $L^2(\mathbb{R}^2)$, since the Gaussian is indefinite. The hidden $su(2)$ algebra uncovered here relies crucially on the presence of a
	nontrivial nilpotent operator associated with the resonant (Jordan-type)
	structure of the degenerate PU oscillator. In the non-degenerate case the
	Hamiltonian is fully diagonalisable into two independent oscillators and no such
	nilpotent mixing operator exists, so this algebraic structure is genuinely
	absent away from the degenerate point.
	
	It is instructive to contrast this non-diagonalisable realisation with the
	simpler Hamiltonian
	\begin{equation}
	H_2 := \sqrt{2}\,\kappa K ,  \label{secondH}
	\end{equation}
	which arises naturally in the next section as a bi-Hamiltonian partner to $H_g$, giving rise to the same classical flow. Since in the quantum theory $K$ lies in the centre of the
	representation, $H_2$ commutes with all $su(2)$ generators and is fully
	diagonalisable. In this case the same $su(2)$ algebra acts in the conventional
	way as a genuine degeneracy algebra: for fixed eigenvalue $k$, all states in
	the finite-dimensional multiplet $\mathcal H_k$ share the same energy
	$E^{(2)}_k=\sqrt{2}\,\kappa k$, and the ladder operators $M_\pm$ generate an
	$(k+1)$-fold degenerate eigenspace.
	
	This provides a sharp contrast with the ghostly Hamiltonian $H_g$, for which the
	term $M_-$ breaks the degeneracy and renders the action non-diagonalisable. The
	same $su(2)$ algebra thus interpolates between two qualitatively different
	roles: for $H_2$ it generates true energy degeneracies, while for $H_g$ it
	organises finite Jordan chains associated with a single eigenvalue.

\section{Bi-Hamiltonian structure and quantisation ambiguity}	
Following \cite{FFT,felski2026three}, we may attempt to exploit the bi-Hamiltonian structure to identify a positive-definite Hamiltonian in a suitable parameter regime that admits normalisable eigenfunctions.

Defining the vector $\vec{x} = \{ x,y,p_x,p_y    \}$ we can generate the classical dynamical system (\ref{flow1}), (\ref{flow2}) with a standard canonical Poisson tensor $J_g$, but alternatively also from a second Hamiltonian $H_2$ with different noncanonical  Poisson tensor $J_2$
\begin{equation}
	\frac{d \vec{x}}{dt} = J_g \cdot \nabla H_g = J_2 \cdot \nabla H_2, \
\end{equation}
where
\begin{eqnarray}
	H_2 &=& \frac{1}{2 \sqrt{2}} \left[ \frac{\nu ^2-3 \Omega  }{\nu ^2-\Omega } p_x^2 - 2 \frac{ \nu ^2+\Omega  }{\nu ^2-\Omega } p_x p_y +\frac{\Omega -3 \nu ^2
		}{\nu ^2-\Omega }  p_y^2   +\frac{1}{2}  \left(3 \nu ^2-\Omega  \right)x^2 	-\left(\nu ^2+\Omega \right) x y\right.     \qquad \\ &&
	\left.   \qquad \qquad \qquad \qquad \qquad \qquad \qquad \qquad  \qquad \qquad \qquad \qquad \qquad \qquad+\frac{1}{2}  \left(3 \Omega -\nu ^2\right) y^2  \right] \notag 
\end{eqnarray}
and 
\begin{equation}
 J_g= \left(
	\begin{array}{cccc}
		0 & 0 & 1 & 0 \\
		0 & 0 & 0 & 1 \\
		-1 & 0 & 0 & 0 \\
		0 & -1 & 0 & 0 
	\end{array}
	\right) ,  \quad
	J_2= \frac{ 1  }{\sqrt{2} \left(\nu ^2-\Omega \right)}  \left(
		\begin{array}{cccc}
			0 & 0 & 3 \nu ^2-\Omega  & -\nu ^2-\Omega  \\
			0 & 0 & \nu ^2+\Omega  & \nu ^2-3 \Omega  \\
			\Omega -3 \nu ^2 & -\nu ^2-\Omega  & 0 & 0 \\
			\nu ^2+\Omega  & 3 \Omega -\nu ^2 & 0 & 0 
		\end{array}
		\right).
\end{equation}
Notice that as an operator, we can represent this second Hamiltonian simply as stated in (\ref{secondH}). Similarly as for the nondegenerate case \cite{FFT}, we may try to form a more general Hamiltonian and Poisson tensor to generate the same flow. Defining the linear combinations
\begin{equation}
       \bar{J} = c_1 J_g + c_2 J_2, \qquad \bar{H} = c_3 H_g + c_4 H_2,
\end{equation}
we obtain the identical flow as in (\ref{flow1}), (\ref{flow2}) when taking
\begin{equation}
	c_3 = \frac{c_1}{\Delta}, \quad c_4 = \frac{2 c_2}{ \Delta}, \quad \text{with}\,\,
	\Delta = c_1^2 + 2 \sqrt{2} c_1 c_2 + 2 c_2^2, \quad c_1 \neq - \sqrt{2} c_2  ,   \label{ccc}
\end{equation}
with $c_1$ and $c_2$ left entirely arbitrary. Recalling that $M_-$ is positive definite an attractive choice would be $c_4= - \sqrt{2} c_3$, as it eliminates the operator $K$ from $\bar{H}$. However, precisely that choice is prevented by the last condition in (\ref{ccc}).

With $c_3$ and $c_3$ as in (\ref{ccc}), we can express $ \bar{H}$ as
\begin{equation}
	\bar{H} = \vec{p}^\intercal   M_p  \vec{p}  + \vec{x}^\intercal   M_v  \vec{x} , \quad \text{with}\,\,
   \vec{p} = (p_x,p_y)^\intercal , \,\,  \vec{x} = (x,y)^\intercal .
\end{equation}
with
{\small
\begin{eqnarray}
M_p&=& \frac{1}{\sqrt{2} (\nu^2 - \Omega) \Delta }	\left(
	\begin{array}{cc}
		\sqrt{2}  \left(\nu ^2-\Omega \right) c_1 + \left(\nu ^2-3 \Omega \right)c_2 & - \left(\nu ^2+\Omega \right) c_2\\
		- \left(\nu ^2+\Omega \right) c_2& \sqrt{2}  \left(\Omega -\nu ^2\right) c_1+\left(\Omega -3 \nu ^2\right) c_2 
	\end{array}
	\right) ,  \qquad\\
	M_v&=& \frac{1}{4 \Delta} \left(
	\begin{array}{cc}
		4 \nu ^2  c_1 +\sqrt{2}  \left(3 \nu ^2-\Omega \right) c_2& -\left(2 c_1+\sqrt{2} c_2\right) \left(\nu ^2+\Omega \right) \\
		-\left(2 c_1+\sqrt{2} c_2\right) \left(\nu ^2+\Omega \right) & 4  \Omega c_1 -\sqrt{2} \left(\nu ^2-3 \Omega \right) c_2 
	\end{array}
	\right) .
\end{eqnarray}}
Despite the fact that $c_1$ and $c_2$ are free parameters, there exists no choice for which $M_p$ and $M_v$ are simultaneously positive definite in any parameter regime, as can be seen from their eigenvalues
{\footnotesize
\begin{eqnarray}
	E_p &=&-\frac{c_2 \left(\nu ^2+\Omega \right)\pm\sqrt{2 c_1^2 \left(\nu ^2-\Omega \right)^2+4 \sqrt{2} c_1 c_2 \left(\nu ^2-\Omega \right)^2+c_2^2 \left(5 \nu ^4-6 \nu
			^2 \Omega +5 \Omega ^2\right)}}{\sqrt{2} \Delta  \left(\nu ^2-\Omega \right)} ,\\
		E_v &=&  \frac{\left(2 c_1+\sqrt{2} c_2\right) \left(\nu ^2+\Omega \right)\pm\sqrt{8 c_1^2 \left(\nu ^4+\Omega ^2\right)+4 \sqrt{2} c_2 c_1 \left(3 \nu ^4-2 \nu ^2 \Omega
				+3 \Omega ^2\right)+2 c_2^2 \left(5 \nu ^4-6 \nu ^2 \Omega +5 \Omega ^2\right)}}{4 \Delta }. \notag
\end{eqnarray} }
In \cite{kaparulin2020resonance} a further conserved quantity was reported
\begin{equation}
	Q= \omega ^2 \left(q+\frac{  \ddot{q} }{\omega ^2}\right)^2+\left( \dot{q} +\frac{   \dddot{q} }{\omega ^2}\right)^2,
\end{equation}
which transforms to
\begin{equation}
	Q=   \frac{\nu ^2+\Omega  }{\nu ^2-\Omega
}     \left[  \frac{\left(p_x-p_y\right){}^2}{\nu ^2-\Omega}+  \frac{1}{2} (x-y)^2 \right]  .
\end{equation} 
However, despite being conserved in time, it does not admit the interpretation as a Hamiltonian, as does not lead to the PU-equation of motion.

Our overall conclusion is that, unlike as in the nondegenerate model \cite{FFT}, the bi-Hamiltonian structure cannot be utilised to identify a positive-definite Hamiltonian that generates the system’s flow in the degenerate model.

\section{Factorisation and effective one-dimensional Hamiltonians}

In this section we explore whether any remnant of a physically acceptable
Hilbert-space structure survives at the resonant point, despite the
non-normalisability of the full ghostly spectrum. Although the formal ground
state (\ref{wansatz}) is not square integrable in $L^2(\mathbb R^2)$, it turns out
that its Gaussian form can be factorised into two one-dimensional components,
one of which may be normalisable in a restricted parameter regime. This
suggests the possibility of isolating an effective one-dimensional sector with
a well-defined quantum dynamics, even though the full two-dimensional theory
remains ghostly.

To this end, we write the ``ground state'' in (\ref{wansatz}) in the quadratic
form 
\begin{equation}
	\psi_0(x,y) = c_0 e^{-\frac{1}{2} \vec{x}^\intercal   M  \vec{x} } , \qquad   \text{with} \,\, M= \left(
	\begin{array}{cc}
		\alpha_\epsilon^\eta  & -\gamma_\epsilon^\eta  \\
		-\gamma_\epsilon^\eta  & \beta_\epsilon^\eta  \\
	\end{array}
	\right),  \qquad    \vec{x} = (x,y)^\intercal  ,\label{wansatzfac}
\end{equation}
we diagonalise the quadratic form by introducing the normalised eigenvectors
$v_\pm$ of $M$
\begin{equation}
       M v_\pm  = \lambda_\pm  v_\pm, \quad \lambda_\pm= \frac{1}{2} \left(\alpha +\beta \pm \sqrt{(\alpha -\beta )^2+4 \gamma  ^2}\right), \quad   v_\pm =\left( \frac{ \mp \sqrt{2} \gamma \rho_\pm  }{\lambda_+ -\lambda_-} , \frac{1}{\sqrt{2} \rho_\pm }         \right),
\end{equation}
with
\begin{equation}
	\rho_\pm   = \sqrt{\frac{\lambda _+-\lambda _-}{\pm (\beta -\alpha )+\lambda _+-\lambda _-}}  ,
\end{equation}
to define the unitary matrix
\begin{equation}
U   =        \left(  v_+ v_-      \right)^\intercal  .
\end{equation}
Introducing the variables
\begin{equation}
	\vec{\tilde{x}} =  U \vec{x}, \qquad   \vec{\tilde{x}} = (\tilde{x}_+,\tilde{x}_-)^\intercal, \quad 
	\tilde{x}_\pm  =  \pm \frac{\sqrt{2}  \gamma \rho_\pm }{\lambda_- - \lambda_+}   x + \frac{1}{\sqrt{2} \rho_\pm } y,
\end{equation}
we can factorise the ``ground state" wavefunction as 
\begin{equation}
	\psi_0(x,y) = \phi_0^+(\tilde{x}_+)  \phi_0^-(\tilde{x}_-), \quad \text{with} \,\,   \phi_0^\pm(\tilde{x}_\pm) = c_\pm e^{ - \frac{1}{2}  \lambda_\pm \tilde{x}_\pm     }  .
\end{equation}
This factorisation separates the indefinite Gaussian into two one-dimensional
modes with weights controlled by the eigenvalues $\lambda_\pm$.

The Hamiltonian (\ref{Hghost}) then acquires the form
\begin{eqnarray}
   H &=& \frac{\alpha - \beta}{\lambda_+ - \lambda_-}  \left(p_{\tilde{x}_+}^2 -p_{\tilde{x}_-}^2 \right)-\frac{2}{ \rho_- \rho_+ }p_{\tilde{x}_+} p_{\tilde{x}_-}+\frac{\left(\nu ^2-\Omega \right) (\alpha -\beta )-2 g \gamma (\epsilon ,\eta )}{2 \left(\lambda _+-\lambda _-\right)} \left( \tilde{x}_+^2-\tilde{x}_-^2  \right) \qquad \\
   &&  +	\frac{\nu^2 + \Omega}{2} \left( \tilde{x}_+^2+\tilde{x}_-^2  \right) + \frac{g (\beta -\alpha )+2 \left(\Omega -\nu ^2\right) \gamma }{2 \gamma  \rho _+ \rho _-}  \tilde{x}_+ \tilde{x}_-  . \notag
\end{eqnarray}
It turns out that $\lambda_-$ is always negative for $\epsilon=1$, $\eta=\pm 1$,
so that the corresponding factor $\phi_0^-(\tilde{x}_-)$ is never square
integrable. By contrast, $\lambda_+$ admits regions in parameter space where it
is positive, and in this regime the factor $\phi_0^+(\tilde{x}_+)$ defines a
normalisable one-dimensional Gaussian.

In this regime it is natural to ask whether the normalisable factor
$\phi_0^+(\tilde{x}_+)$ can be used to define an effective one-dimensional
quantum dynamics for the remaining variable $\tilde{x}_-$ by integrating out
the $\tilde{x}_+$ degree of freedom.
\begin{equation}
	H_{\text{eff}}(\tilde{x}_- ) := \int_{-\infty}^{\infty} d\tilde{x}_+  (\phi_0^+)^*(\tilde{x}_+) \left[ H(\tilde{x}_+,\tilde{x}_-) \left(  \phi_0^+(\tilde{x}_+)  \phi_0^-(\tilde{x}_-)     \right)     \right]
\end{equation}
This leads to effective Hamiltonians of the general form
\begin{equation}
	H_{\text{eff}}(\tilde{x}_- ) :=  (a_1 p_{\tilde{x}_-}^2 + a_2 \tilde{x}_-^2 + a_3) \phi_0^-(\tilde{x}_-)
\end{equation}
For the sector of interest here, $\epsilon=1$, $\eta=1$, we find $a_1<0$ and $a_2<0$ throughout the
entire parameter regime. The effective Hamiltonian therefore describes a
harmonic oscillator with an overall negative sign, and hence still carries a
ghost degree of freedom. Thus, although a partially normalisable sector can be
isolated, this factorisation does not by itself lead to a fully physical
one-dimensional quantum theory.

This analysis shows that, although a partial factorisation can isolate a
normalisable direction in a restricted parameter regime, the ghost problem
persists even at the level of the effective one-dimensional dynamics.

\section{Conclusions}

We have analysed the quantum degenerate PU oscillator
from several complementary perspectives, with particular emphasis on its
algebraic organisation and on the ambiguities of quantisation that arise at the
resonant point. At the classical level, the coalescence of the two frequencies
leads to a non-diagonalisable dynamics characterised by secular solutions and a
Jordan-type structure of the time-evolution operator. This singular behaviour
already signals that the degenerate model is not a smooth continuation of the
non-degenerate theory, but a genuinely distinct dynamical system.

At the quantum level, working in the ghostly two-dimensional formulation, we
have shown that the collapse of the standard Fock-space construction is
accompanied by the emergence of a hidden spectrum-generating $su(2)$ algebra.
In contrast to the usual role of such algebras as degeneracy symmetries, the
present algebra organises the generalised eigenvectors of the ghostly
Hamiltonian into finite Jordan chains associated with a single eigenvalue. This
provides a concrete and explicit realisation of how non-diagonalisable quantum
dynamics can be organised algebraically at resonance.

A central conceptual outcome of our analysis is the inequivalence of different
quantisations of classically equivalent Hamiltonians. Although the ghostly
Hamiltonian $H_g$ and the alternative Hamiltonian $H_2$
generate exactly the same classical phase-space flow, their quantisations lead
to radically different quantum theories: the former is non-diagonalisable and
organised by Jordan chains, while the latter is fully diagonalisable and
exhibits genuine degeneracies. The resonant PU oscillator thus provides a
particularly transparent example of the fact that classical equivalence does
not guarantee quantum equivalence, and that the choice of Hamiltonian is an
intrinsic part of the quantisation problem in higher-derivative systems.

Finally, our analysis of the bi-Hamiltonian structure and of the partial
factorisation of the ground state shows that, unlike as in the non-degenerate
model, no positive-definite Hamiltonian can be constructed that generates the
full degenerate dynamics, although restricted ghost-free sectors may still
exist in limited parameter regimes. These results reinforce the view that the
resonant PU oscillator occupies a singular position in the space of
higher-derivative theories, and provides a good testing ground for proposed
resolutions of the ghost problem and for the limits of algebraic and geometric
quantisation methods in higher-time-derivative systems.

We emphasise that, despite the algebraic control achieved in this work, the
resulting quantum theory is still not a fully satisfactory Hilbert-space
quantisation. In particular, for real $\nu^2>\Omega$ the formal eigenfunctions of
the ghostly Hamiltonian remain non-normalisable in $L^2(\mathbb R^2)$, reflecting
the indefinite Gaussian weight and the underlying ghost nature of the model.
The hidden $su(2)$ algebra and the organisation into Jordan chains therefore do
not by themselves resolve the problem of defining a physical inner product in the surviving sector at the resonant point.

A natural direction for future work is to investigate whether a suitable
non-unitary similarity transformation can map the present ghostly formulation
to an equivalent representation with a positive-definite metric and
normalisable eigenstates, in the spirit of recent developments in
ghost-free quantisation of HTDTs
\cite{fring2025ghost}. Such a construction would provide a decisive test of
whether the resonant PU oscillator admits a fully consistent
quantum realisation beyond the formal algebraic level. It would also be very interesting to investigate whether the kind of hidden
spectrum-organising Lie-algebraic structures uncovered here persist in ghostly
field-theoretic systems, such as the $(1+1)$-dimensional models with ghostly
interactions discussed in \cite{deffayet2025ghostly}, and whether they play a
structural role analogous to that found here.

In this sense, the degenerate PU oscillator remains a particularly
fertile testing ground for the interplay between algebraic structures, ghosts, and
the foundations of quantisation in higher-time-derivative systems.

\medskip

\noindent {\bf Acknowledgments}: IM was supported by the Australian Research Council Future Fellowship FT180100099. TT was supported by R7 (2025) Young Researchers Support Project, Faculty of Science, Kyushu University.

\newif\ifabfull\abfulltrue


\begin{thebibliography}{10}
	
	\bibitem{buchbinder2017effective}
	I.~L. Buchbinder,
	\newblock {\em Effective action in quantum gravity},
	\newblock (New York, Routledge), 2017.
	
	\bibitem{stelle77ren}
	K.~S. Stelle,
	\newblock Renormalization of higher-derivative quantum gravity,
	\newblock Phys. Rev. D {\bf 16}(4), 953 (1977).
	
	\bibitem{starobinsky1980new}
	A.~A. Starobinsky,
	\newblock A new type of isotropic cosmological models without singularity,
	\newblock Phys. Lett. B {\bf 91}, 99--102 (1980).
	
	\bibitem{sotiriou2010f}
	T.~P. Sotiriou and V.~Faraoni,
	\newblock $f(R)$ theories of gravity,
	\newblock Rev. Mod. Phys. {\bf 82}, 451--497 (2010).
	
	\bibitem{grav3}
	A.~V. Smilga,
	\newblock Spontaneous generation of the Newton constant in the renormalizable
	gravity theory,
	\newblock ITEP preprint 63 (1982) 8 pp, arXiv preprint arXiv:1406.5613 (2014)
	(1982).
	
	\bibitem{tomboulis2015renormalization}
	E.~Tomboulis,
	\newblock Renormalization and unitarity in higher derivative and nonlocal
	gravity theories,
	\newblock Mod. Phys. Lett. A {\bf 30}, 1540005 (2015).
	
	\bibitem{modesto2012super}
	L.~Modesto,
	\newblock Super-renormalizable quantum gravity,
	\newblock Phys. Rev. D {\bf 86}, 044005 (2012).
	
	\bibitem{modesto2016superrenormalizable}
	L.~Modesto and I.~L. Shapiro,
	\newblock Superrenormalizable quantum gravity with complex ghosts,
	\newblock Phys. Lett. B {\bf 755}, 279--284 (2016).
	
	\bibitem{barnaby2008dynamics}
	N.~Barnaby and N.~Kamran,
	\newblock Dynamics with infinitely many derivatives: the initial value problem,
	\newblock JHEP {\bf 2008}(02), 008 (2008).
	
	\bibitem{ghostconst}
	T.-J. Chen, M.~Fasiello, E.~A. Lim, and A.~J. Tolley,
	\newblock Higher derivative theories with constraints: Exorcising
	Ostrogradski's Ghost,
	\newblock J. Cosmol. Astropart. Phys. {\bf 2013}(02), 042 (2013).
	
	\bibitem{motohashi1}
	H.~Motohashi, K.~Noui, T.~Suyama, M.~Yamaguchi, and D.~Langlois,
	\newblock Healthy degenerate theories with higher derivatives,
	\newblock J. Cosmol. Astropart. Phys. {\bf 2016}(07), 033 (2016).
	
	\bibitem{motohashi4}
	H.~Motohashi and T.~Suyama,
	\newblock Quantum Ostrogradsky theorem,
	\newblock JHEP {\bf 2020}(9), 1--10 (2020).
	
	\bibitem{weldon03quant}
	H.~A. Weldon,
	\newblock Quantization of higher-derivative field theories,
	\newblock Ann. Phys. {\bf 305}(2), 137--150 (2003).
	
	\bibitem{Woodard1}
	R.~P. Woodard,
	\newblock Ostrogradsky's theorem on Hamiltonian instability,
	\newblock Scholarpedia {\bf 10}(8), 32243 (2015).
	
	\bibitem{fring2024higher}
	A.~Fring, T.~Taira, and B.~Turner,
	\newblock Higher Time-Derivative Theories from Space--Time Interchanged
	Integrable Field Theories,
	\newblock Universe {\bf 10}(5), 198 (2024).
	
	\bibitem{fring2025quant}
	A.~Fring, T.~Taira, and B.~Turner,
	\newblock Quantisations of exactly solvable ghostly models,
	\newblock J. Phys. A: Math. Theor. {\bf 58}, 235301 (2025).
	
	
	\bibitem{pais1950field}
	A.~Pais and G.~E. Uhlenbeck,
	\newblock On field theories with non-localized action,
	\newblock Phys. Rev. {\bf 79}, 145 (1950).
	
	\bibitem{bolonek2005ham}
	K.~Bolonek and P.~Kosinski,
	\newblock Hamiltonian Structures for Pais--Uhlenbeck Oscillator,
	\newblock Acta Phys. Polon. B {\bf 36}(6), 2115 (2005).
	
	\bibitem{mann2005dirac}
	P.~D. Mannheim and A.~Davidson,
	\newblock Dirac quantization of the Pais-Uhlenbeck fourth order oscillator,
	\newblock Phys. Rev. A {\bf71}, 042110 (2005).
	
	\bibitem{dam2006}
	E.~Damaskinsky and M.~Sokolov,
	\newblock Remarks on quantization of Pais--Uhlenbeck oscillators,
	\newblock J. Phys. A: Math. Gen. {\bf 39}, 10499 (2006).
	
	\bibitem{bolonek2006comment}
	K.~Bolonek and P.~Kosinski,
	\newblock Comment on" Dirac Quantization of Pais-Uhlenbeck Fourth Order
	Oscillator",
	\newblock arXiv preprint quant-ph/0612009  (2006).
	
	\bibitem{bender2008no}
	C.~M. Bender and P.~D. Mannheim,
	\newblock No-ghost theorem for the fourth-order derivative Pais-Uhlenbeck
	oscillator model,
	\newblock Phys. Rev. Lett. {\bf 100}(11), 110402 (2008).
	
	\bibitem{smilga2009comments}
	A.~V. Smilga,
	\newblock Comments on the dynamics of the Pais-Uhlenbeck oscillator,
	\newblock SIGMA	{\bf 5}, 017 (2009).
	
	
	\bibitem{most2010h}
	A.~Mostafazadeh,
	\newblock A Hamiltonian formulation of the Pais--Uhlenbeck oscillator that
	yields a stable and unitary quantum system,
	\newblock Phys. Lett. A {\bf 375}, 93--98 (2010).
	
	\bibitem{mosta2011im}
	A.~Mostafazadeh,
	\newblock Imaginary-scaling versus indefinite-metric quantization of the
	Pais-Uhlenbeck oscillator,
	\newblock Phys. Rev. D {\bf 84}, 105018 (2011).
	
	
	\bibitem{andrzejewski2014ham}
	K.~Andrzejewski,
	\newblock Hamiltonian formalisms and symmetries of the Pais-Uhlenbeck
	oscillator,
	\newblock Nucl. Phys. B {\bf 889}, 333--350 (2014).
	
	\bibitem{Sugg3}
	M.~Avendano-Camacho, J.~A. Vallejo, and Y.~Vorobiev,
	\newblock A perturbation theory approach to the stability of the Pais-Uhlenbeck
	oscillator,
	\newblock J. Math. Phys. {\bf 58}(9) (2017).
	
	\bibitem{smilga2017class}
	A.~Smilga,
	\newblock Classical and quantum dynamics of higher-derivative systems,
	\newblock Int. J. Mod. Phys. A {\bf 32}(33), 1730025
	(2017).
	
	\bibitem{mandal2022bfv}
	B.~P. Mandal, V.~K. Pandey, and R.~Thibes,
	\newblock BFV quantization and BRST symmetries of the gauge invariant
	fourth-order Pais-Uhlenbeck oscillator,
	\newblock Nucl. Phys. B {\bf 982}, 115905 (2022).
	
	\bibitem{elbistan2023various}
	M.~Elbistan and K.~Andrzejewski,
	\newblock Various disguises of the Pais-Uhlenbeck oscillator,
	\newblock Nucl. Phys. B {\bf 994}, 116327 (2023).
	
	
\bibitem{FFT}
A.~Felski, A.~Fring, and B.~Turner,
\newblock Lie symmetries and ghost-free representations of the Pais--Uhlenbeck
model,
\newblock Mod. Phys. Lett. A, published online (2025),
\newblock doi:10.1142/S0217732326500197,
\newblock arXiv:2505.07869 [hep-th].
	
	\bibitem{diez2024foundations}
	V.~E. D{\'\i}ez, J.~G. Gaset~Rif{\`a}, and G.~Staudt,
	\newblock Foundations of Ghost Stability,
	\newblock Fortschr. Phys., 2400268 (2024).
	
	\bibitem{felski2026three}
	A.~Felski, A.~Fring, and B.~Turner,
	\newblock Three-dimensional ghost-free representations of the Pais-Uhlenbeck
	model from Tri-Hamiltonians,
	\newblock Phys. Lett. A , 131332 (2026).
	
	\bibitem{kaparulin2020resonance}
	D.~S. Kaparulin, S.~L. Lyakhovich, and O.~D. Nosyrev,
	\newblock Resonance and stability of higher derivative theories of a derived
	type,
	\newblock Phys. Rev. D {\bf 101}, 125004 (2020).
	
	\bibitem{Witten:1981nf}
	E.~Witten,
	\newblock Dynamical breaking of supersymmetry,
	\newblock Nucl. Phys. {\bf B188}, 513 (1981).
	
	\bibitem{Cooper:1982dm}
	F.~Cooper and B.~Freedman,
	\newblock {Aspects of Supersymmetric Quantum Mechanics},
	\newblock Ann. Phys. {\bf 146}, 262 (1983).
	
	\bibitem{bagchi2000super}
	B.~K. Bagchi,
	\newblock {\em Supersymmetry in quantum and classical mechanics},
	\newblock CRC Press, 2000.
	
	\bibitem{darboux}
	G.~Darboux,
	\newblock On a proposition relative to linear equations,
	\newblock C. R. Acad. Sci. Paris {\bf 94}, 1456--1459 (1882),
	\newblock arXiv:physics/9908003.
	
	
	\bibitem{crum}
	M.~M. Crum,
	\newblock Associated Sturm-Liouville systems,
	\newblock Q. J. Math. {\bf 6}, 121--127 (1955).
	
	\bibitem{matveevdarboux}
	V.~B. Matveev and M.~A. Salle,
	\newblock Darboux transformation and solitons,
	\newblock (Springer, Berlin)  (1991).
	
	\bibitem{cannata2010exactly}
	F.~Cannata, M.~Ioffe, and D.~Nishnianidze,
	\newblock Exactly solvable nonseparable and nondiagonalizable two-dimensional
	model with quadratic complex interaction,
	\newblock J. Math. Phys. {\bf 51}(2) (2010).
	
	\bibitem{marquette22lad1}
	I.~Marquette and C.~Quesne,
	\newblock Ladder Operators and Hidden Algebras for Shape Invariant Nonseparable
	and Nondiagonalizable Modelswith Quadratic Complex Interaction. I.
	Two-Dimensional Model,
	\newblock SIGMA	{\bf 18}, 004 (2022).
	
	\bibitem{marquette22lad2}
	I.~Marquette and C.~Quesne,
	\newblock Ladder Operators and Hidden Algebras for Shape Invariant Nonseparable
	and Nondiagonalizable Modelswith Quadratic Complex Interaction. II.
	Three-Dimensional Model,
	\newblock SIGMA	{\bf 18}, 005 (2022).
	
	\bibitem{fring2025ghost}
	A.~Fring, T.~Taira, and B.~Turner,
	\newblock Ghost-Free Quantisation of Higher Time-Derivative Theories via
	Non-Unitary Similarity Transformations,
	\newblock preprint arXiv:2506.21400  (2025).

\bibitem{deffayet2025ghostly}
C.~Deffayet, A.~Held, S.~Mukohyama, and A.~Vikman,
\newblock Ghostly interactions in $(1+1)$-dimensional classical field theory,
\newblock Phys. Rev. D {\bf 112}, 065011 (2025).
	
\end{thebibliography}
\end{document}